%% file: paper.tex
\documentclass{iaus}
\usepackage{graphicx}
\usepackage{subfigure}
\usepackage{amsmath}
\usepackage{natbib}
\usepackage{bm}
\graphicspath{{./fig/}{./png/}}
\input{macros}

\title[Turbulent magnetic pressure instability]
{Turbulent magnetic pressure instability in stratified turbulence}

\author[K. Kemel et al.]
{K. Kemel$^{1,2}$, A. Brandenburg$^{1,2}$,
N. Kleeorin$^3$, and I. Rogachevskii$^3$}

\affiliation{$^1$NORDITA, AlbaNova University Center, Roslagstullsbacken 23, SE-10691 Stockholm, Sweden \\
$^2$Department of Astronomy, Stockholm University, SE--10691 Stockholm, Sweden\\
$^3$Department of Mechanical Engineering, Ben-Gurion University of the Negev,\\
POB 653, Beer-Sheva 84105, Israel}

\pubyear{2010}
\volume{274}
\pagerange{}
\jname{Advances in Plasma Astrophysics}
\editors{}

\begin{document}

\maketitle

\begin{abstract}

A reduction of total mean turbulent pressure due to the presence of magnetic fields
was previously shown to be a measurable effect in direct numerical simulations.
However, in the studied parameter regime the formation of large-scale structures, 
as anticipated from earlier mean-field simulations, was not found.
An analysis of the relevant mean-field parameter dependency and the
parameter domain of interest is conducted 
in order to clarify this apparent discrepancy. 

\end{abstract}

Strong magnetic fields at the solar surface are generally thought to originate from 
the coherent rise of magnetic flux tubes from the tachocline through the solar convection zone.
While the idea is elegant, the question remains whether it can be considered as more than a toy model, 
as the physics of the creation and rise of these flux tubes is not sufficiently understood \citep{Par09}.
Thus far any `successful' numerical simulation of this process had to rely on strong
assumptions, be it in the initial conditions or in simplified equations such
as the thin flux tube approximation \citep{Spr81}.
As such it makes sense to explore alternative mechanisms of magnetic structure formation.
Several models have been proposed where large-scale magnetic field concentrations are created 
through instabilities at the solar surface \citep[][hereafter BKR]{KM00,BKR10}.

Turbulence is generally associated with enhanced transport effects. However, 
it can also generate structures on much larger scales than its driving scales;
see as an example the inverse cascade in 2D hydrodynamic turbulence \citep{Kra67}
or 3D MHD turbulence with magnetic helicity \citep{Fri75}.
%
%
We study here the interaction of the turbulence with
a background magnetic field. 
From the approximate conservation of total turbulent energy $E_{\rm tot}$,
BKR find a reversed feedback from the magnetic fluctuations
on the turbulent pressure \citep[][hereafter RK]{RK07}:
\begin{displaymath}
P_{\rm turb} = -\textstyle{\frac{1}{6}}\overline{b^2}/\mu_0
+\textstyle{\frac{2}{3}}E_{\rm tot}.
\label{Pturb}
\end{displaymath}
It can be seen that the  effective mean magnetic pressure force is reduced and 
can be reversed in a certain parameter range.
It was suggested by RK that this positive feedback could lead to an instability,
resulting in the concentration of magnetic flux.

\begin{figure}\begin{center}
\includegraphics[width=0.54\textwidth]{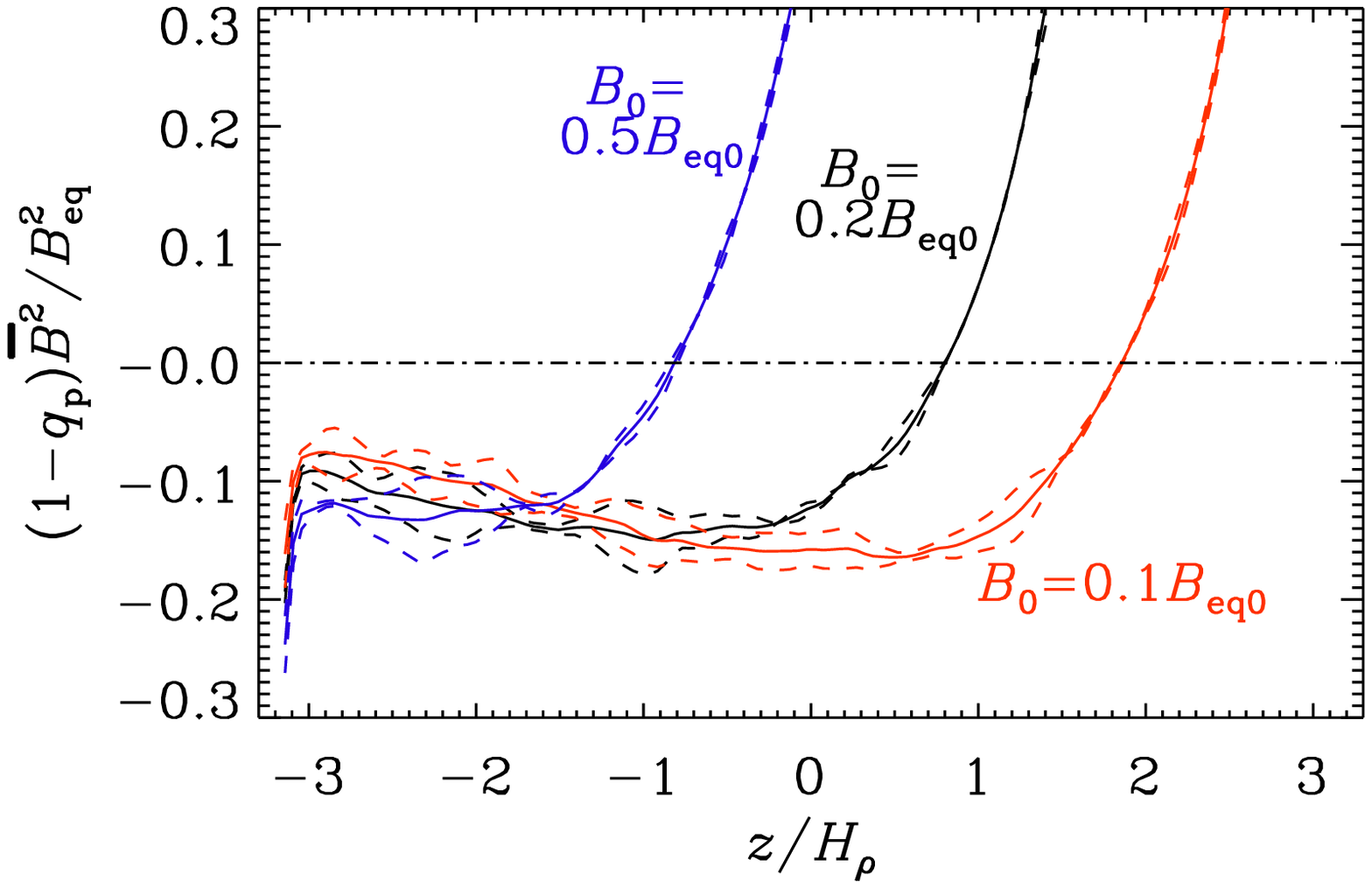}
\includegraphics[width=0.44\textwidth]{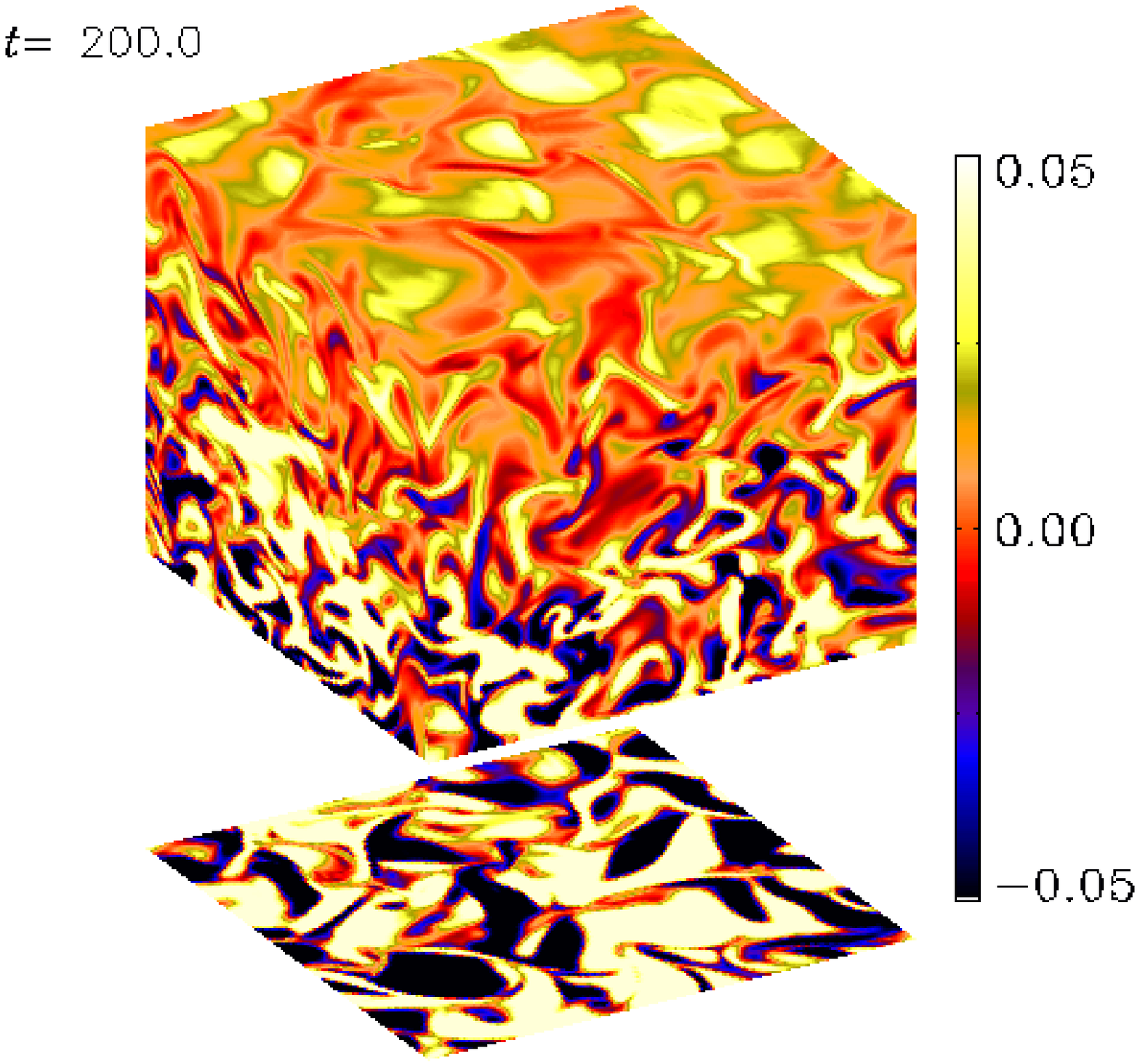}
\end{center}\caption[]{
{\it Left}: Normalized effective mean magnetic pressure as a function of depth for $B_0=0.1 B_{\rm eq0}$, $B_0=0.2 B_{\rm eq0}$, and $B_0=0.5 B_{\rm eq0}$ using $\Rey=120$ and $\Pm=1$. Adapted from \cite{BKKR10}.
{\it Right}: Visualization of $B_y-B_0$ on the periphery of the computational domain $B_0=0.1 B_{\rm eq0}$, and $\Pm=2$. Adapted from \cite{KBKR10}.
}\label{pxyaver_compp}
\end{figure}

BKR confirmed the validity of approximate turbulent energy conservation
using direct numerical simulations (DNS)
of homogeneous isothermal turbulence and they also demonstrated
the basic phenomenon of magnetic flux concentration through the 
interaction between turbulence and the mean Lorentz force in mean-field MHD simulations
which led to a linear instability for sufficiently strong stratification.
%
%
These results were recently corroborated using DNS of
inhomogeneous (stratified) isothermal turbulence of a cubic computational domain.
An example is shown in \Fig{pxyaver_compp} where we show that
the normalized effective mean magnetic pressure
does indeed become negative in a large part
of the domain for different imposed field strengths. 
However, the formation of large-scale structures is not observed,
as can be seen from the right-hand panel  of \Fig{pxyaver_compp}.

It is important to understand whether there  is really a conflict
between DNS and mean-field results, or if these two studies
simply apply to different parameter regimes.
In order to have a chance to resolve this question, we conduct a
systematic parameter survey of the instability in the mean-field model.
More specifically, we determine the functional dependence of the
growth rate on the input variables in order to find the relevant 
parameter space for the instability to develop.

In an isothermal stratified box we solve the equations in two dimensions:
\EQ
{\partial\meanUU\over\partial t}=-\meanUU\cdot\nab\meanUU
-\cs^2\nab\ln\meanrho+\grav+\meanFFFF_{\rm M}+\meanFFFF_{\rm K,tot},
\nonumber
\EN
where
\EQ
\meanrho \, \meanFFFF_{\rm M} = -\half\nab[(1-q_{\rm p})\meanBB^2]
+ \meanBB \cdot \nab\left[(1-q_{\rm s})\meanBB\right]
\nonumber
\EN
is the mean-field Lorentz force and
\EQ
\meanFFFF_{\rm K,tot}=(\nut+\nu)\left(\nabla^2\meanUU+\nab\nab\cdot\meanUU
+2\meanSSSS\nab\ln\meanrho\right)
\nonumber
\EN
is the total (turbulent and microscopic) viscous force with
$\SSSS$ being the viscous stress tensor.
In addition, we solve the continuity and uncurled induction equations,
\EQ
{\partial\ln\meanrho\over\partial t}=-\meanUU\cdot\nab\ln\meanrho
-\nab\cdot\meanUU,\quad
{\partial\meanAA\over\partial t}=\meanUU\times\meanBB-(\etat+\eta)\meanJJ.
\nonumber
\EN
We adopt a Cartesian coordinate system, $(x,y,z)$.
The mean field is given by $\meanBB=(0,B_0,0)+\nab\times\meanAA$,
and the vertical gravitational acceleration is $\grav=(0,0,-g)$.
The other input parameters of the simulations are the sound speed $c_{\rm s}$,
the density at the top of the box $\rho_{\rm top}$, 
the pressure coefficient $q_{\rm p}$ (RK), the magnetic Prandtl number $\Pm$
and the molecular diffusivity $\eta$.

\begin{figure}\begin{center}

\includegraphics[width=1.\textwidth]{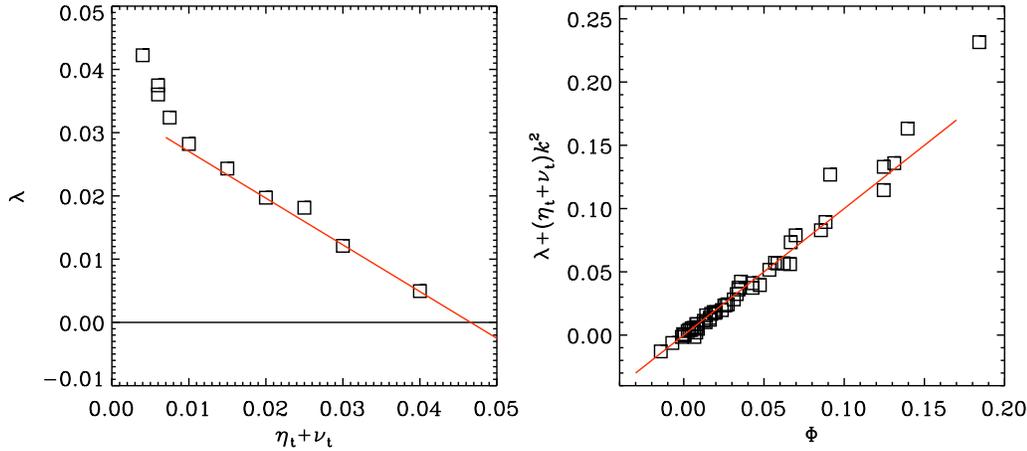}
\end{center}\caption[]{
{\it Left}: dependence of growth rate on turbulent diffusivity and viscosity. 
{\it Right}: modified growth rate as a function of $\Phi$.
}\label{lamfit_full8b.ps}
\end{figure}

We start by isolating the dependence of the growth rate on the turbulent diffusivity 
and find approximately linear behaviour. 
From a data survey we observe, depending on the applied field, 
different diffusivity values above which  the instability does not develop,
which would imply
\EQ
\lambda=\Phi(B)-(\etat+\nut)k^2
\nonumber
\EN
where $1/k$ is a length scale introduced for dimensional reasons.
We find that $\Phi$ is proportional to the  Alfv\'en speed at the top
of the box, where the instability initiates.
Eventually we arrive at the fit formula
\EQ
\lambda+\left(1+\Pmt\right)\etat k^2
=\vA k_x\left(1+q_{\rm p}/q_{\rm p}^*\right)\exp\left(\ell_z/H_\rho\right),
\nonumber
\EN
as shown in \Fig{lamfit_full8b.ps}. 
Here, $\Pmt=\nut/\etat$, $\ell_z$ is some typical vertical length scale,
and $H_\rho=\cs^2/g$ is the density scale height.
The instability resulting from this feedback effect
was verified but not observed to generate large-scale structures in DNS
with the current scale separation and parameter range studied so far.

An extension of this work will be the inclusion of more physics
(e.g.\ radiative cooling) in the DNS 
and the comparison with mean-field models in a 3D setup.
As was already pointed out in BKR, the mean-field model predicts
additional short-wavelength perturbutions along the direction of the
mean magnetic field, that were not included in the present study.




\end{document}

%% file: macros.tex
\newcommand{\EQ}{\begin{equation}}
\newcommand{\EN}{\end{equation}}
\newcommand{\EQA}{\begin{eqnarray}}
\newcommand{\ENA}{\end{eqnarray}}

\newcommand{\Fig}[1]{Figure~\ref{#1}}

\newcommand{\meanrho}{\overline{\rho}}

{}
{}
\newcommand{\meanFFFF}{\overline{\mbox{\boldmath ${\cal F}$}}{}}{}

\newcommand{\meanSSSS}{\overline{\mbox{\boldmath ${\mathsf S}$}} {}}

{}
{}
{}
{}
{}
{}
{}
{}
\newcommand{\meanAA}{\overline{\mbox{\boldmath $A$}}{}}{}
\newcommand{\meanBB}{\overline{\mbox{\boldmath $B$}}{}}{}
{}
{}
{}
{}
{}
{}
{}
{}
\newcommand{\meanJJ}{\overline{\mbox{\boldmath $J$}}{}}{}
{}
\newcommand{\meanUU}{\overline{\bm{U}}}

{}
{}
{}

{}

{}
{}

\newcommand{\grav}{\mbox{\boldmath $g$} {}}
\newcommand{\nab}{\mbox{\boldmath $\nabla$} {}}

\newcommand{\SSSS}{\mbox{\boldmath ${\sf S}$} {}}

\def\Pm{\mbox{\rm Pr}_M}
\def\Pmt{\mbox{\rm Pr}_M^{\rm turb}}

\def\Rey{\mbox{\rm Re}}

\def\cs{c_{\rm s}}

\def\vA{v_{\rm A}}

\def\nut{\nu_{\rm t}}
\def\etat{\eta_{\rm t}}

\def\half{{\textstyle{1\over2}}}

\newcommand{\yssr}[3]{ #1, {\textit Spa.\ Sci.\ Rev.,} {#2}, #3}

\newcommand{\yan}[3]{ #1, \textit{Astron.\ Nachr.,} \textit{#2}, #3}

\newcommand{\yana}[3]{ #1, \textit{A\&A,} \textit{#2}, #3}

\newcommand{\yjfm}[3]{ #1, \textit{J.\ Fluid Mech.,} \textit{#2}, #3}

\newcommand{\ypf}[3]{ #1, \textit{Phys.\ Fluids,} \textit{#2}, #3}

\newcommand{\ysph}[3]{ #1, \textit{Solar Phys.,} \textit{#2}, #3}

\newcommand{\ypre}[3]{ #1, \textit{Phys.\ Rev.\ E,} \textit{#2}, #3}

\newcommand{\yjour}[4]{ #1, \textit{#2}, \textit{#3}, #4}
